\documentclass[fleqn,usenatbib]{mnras}

\usepackage{newtxtext,newtxmath}

\usepackage[T1]{fontenc}

\DeclareRobustCommand{\VAN}[3]{#2}
\let\VANthebibliography\thebibliography
\def\thebibliography{\DeclareRobustCommand{\VAN}[3]{##3}\VANthebibliography}

\usepackage{graphicx}	
\usepackage{amsmath}	
\usepackage{booktabs}

\title{New Transient ULX Candidate in NGC 4254: \\ Evidence of Circumbinary Disk?}

\author[S. Allak]{Sinan Allak$^{1}$\thanks{E-mail:0417allaksinan@gmail.com} \\
$^1$Space Science and Solar Energy Research and Application Center (UZAYMER), University of Çukurova, 01330, Adana, Turkey\\
}

\date{Accepted 2023 October 12. Received 2023 October 12; in original form 2023 September 07}

\pubyear{2023}

\begin{document}
\label{firstpage}
\pagerange{\pageref{firstpage}--\pageref{lastpage}}
\maketitle

\begin{abstract}
This paper presents the identification of a new transient ULX candidate (ULX-3) with reaching a peak luminosity of $\sim$ 4 $\times$ 10$^{39}$ erg s$^{-1}$ in NGC 4254 by using archival {\it Chandra}, {\it Swift X-Ray Telescope (Swift/XRT)}, {\it Hubble Space Telescope (HST)}, and {\it James Webb Space Telescope (JWST)} observations. From precise astrometric calculations, unique optical, near-infrared (NIR) and mid-infrared (mid-IR) counterparts were found. The spectral energy distribution (SED) and color-magnitude diagrams (CMDs) of counterparts of the new ULX candidate were plotted to constrain the nature of the possible donor star. Evidence of a circumbinary disk was found from its SED with two \textit{blackbody} temperatures of 1000 and 200 K. Moreover, according to the X-ray hardness ratios, ULX-3 exhibits very hard to very soft transitions as seen in some high-mass X-ray binaries (HMXBs) with Be-star donors Moreover, ULX-3 varies by more than two orders of magnitude in the 0.3-10 keV energy band as seen in typical transient ULXs.
\end{abstract}
,
\begin{keywords}
galaxies: individual: NGC 4254 - X-rays: binaries (ULXs) -stars: general: (Counterparts of ULXs), -(stars:) circumstellar matter: Circumbinary disk/dust – X-rays: individual (NGC 4254 ULX-3) -space vehicles: instruments: JWST
\end{keywords}

\section{Introduction}

Ultraluminous X-ray sources (ULXs) are usually defined by isotropic fluxes that give X-ray luminosities (L$_{X}$) are $\geq$ 10$^{39}$ erg s$^{-1}$ for accretor masses 10 M$\odot$ \citep{2023NewAR..9601672K}. Early models for ULXs have suggested that the accreting mass was unusually large 10$^{2}$ - 10$^{4}$ M$\odot$ (e.g. \citealp{1999ApJ...519...89C}) that such as these sources accreting at significantly sub-Eddington rates were proposed (e.g. \citealp{2004ApJ...614L.117M}). The discovery of pulsar ULXs (PULXs) has shown that the luminosities of ULXs can be highly super-Eddington. Moreover, the discovery of several PULXs (e.g. \citealp{2014Natur.514..202B,2018MNRAS.476L..45C,2019MNRAS.488L..35S}) and also detecting cyclotron resonance scattering feature for M51 ULX-8 \citep{2018NatAs...2..312B}, it has led to a reinterpretation of the standard super-critical model due to the magnetosphere. The majority of ULXs are relatively persistent sources, but, recent studies have identified a very high number of transient ULXs (e.g. \citealp{2016A&A...593A..16T,2020ApJ...895..127B,2020MNRAS.499.5682A,2021MNRAS.501.1002W,2022MNRAS.510.4355A,2023ApJ...951...51B}). The most distinctive feature of these sources is that they show X-ray variability from an order magnitude to several orders magnitude.

Many observational studies have been carried out to identify the optical counterparts of ULXs and constrain the nature of the optical emission (e.g., \citealp{2011ApJ...737...81T,2012MNRAS.420.3599S,2013ApJS..206...14G,2018ApJ...854..176V,2021MNRAS.505..771E,2022MNRAS.515.3632A,2022MNRAS.517.3495A,2023ApJ...946...72G}). However, it is difficult to distinguish exactly where the optical emission from ULXs comes from since it can be observed from the photosphere of the donor star, the outer accretion disk, the accretion disks, or a combination of these \citep{2011AN....332..398R,2011ApJ...737...81T,2015NatPh..11..551F,2022MNRAS.515.3632A}. On the other hand, the optical spectrum of ULX NGC 7793 P13 has shown the absorption lines that indicate that observed optical emission comes from B9Ia supergiant type star with a mass of 18-23 M$\odot$ \citep{2014Natur.514..198M}. Moreover, optical counterparts have also been reported to exhibit variability in the optical band when sufficient data are available such as transient source ULX-4 M51 \citep{2022MNRAS.510.4355A} and variable sources ULX-8 M51 \citep{2022MNRAS.517.3495A} and NGC 1313 X-2 and NGC 1313 \citep{2009ApJ...690L..39L}. The common feature of these three ULXs is that the observed optical emission source is most likely an accretion disc. The most optical counterparts of ULXs are faint m$_{V_Vega}$ > 18 mag. The observability of optical counterparts is limited since they are usually in crowded stellar groups and located in interstellar and/or local gas and dust.

Infrared (IR) wavelength observations, which are relatively less susceptible to the effects of extinction such as dust, offer valuable constraints for investigating the mass donor stars and the circumstellar environment of ULXs. This can be a good tool to avoid confounding multiple emitting components within ULX systems (\citealp{2019ApJ...878...71L,2016ApJ...831...88D} and references therein). Many ULXs were reported by \citep{2014MNRAS.442.1054H} and \citep{2017MNRAS.469..671L,2020MNRAS.497..917L} with near-IR (NIR) observations with absolute magnitudes consistent with red supergiants (RSGs). Moreover, \cite{2015MNRAS.453.3510H,2016MNRAS.459..771H} presented NIR spectroscopy studying for ULXs and they claimed that three donors of ULXs likely RSG masses. Moreover, by using IR to X-ray observations, \citep{2016ApJ...831...88D,2019ApJ...878...71L} have reported the origin of X-1 Holmberg IX mid-IR emission is either circumstellar dust or a variable jet. On the other hand, the nature of IR emission of ULXs is still unclear, since they are variable, which has been observed in the X-ray on timescales of hours to years (e.g., \citep{2009MNRAS.397.1061H,2017ApJ...838L..17L}) Moreover, since ULXs are located in crowded regions of the host galaxies, {\it JWST} images with good enough resolution are critical to investigate their accurate optical, NIR and mid-NIR counterparts. Most of the point-like and/or bright NIR counterparts of ULXs observed in past studies are likely unresolved sources are blend sources, therefore, many sources may not be red enough for them to be an RSG \citep{2023arXiv230611163A,2023arXiv230800141A}. The significant improvement in resolution provided by {\it JWST} has led to a remarkable new perspective on the unclear nature of ULX donors, necessitating a major reassessment of previous IR image investigations into counterparts of ULXs. \\

This work is carried out as an X-ray to the multiwavelength study of the new transient ULX candidate (ULX-3) in NGC 4254 observed with {\it Swift-XRT}, {\it Chandra}, {\it JWST} and {\it HST}. One of the primary goals of this work is to search for and identify possible optical, NIR, and mid-IR counterparts of this new ULX candidate by precise astrometric calculations. The IR emission of the possible donor star is investigated in detail to constrain main properties such as spectral type and mass as well as the source of the emission. Moreover, the X-ray spectral and temporal properties of the new ULX candidate are analyzed in detail, aiming to impose constraints on the compact object nature as well as the mechanism of X-ray emission. The paper is structured as follows. Section \ref{sec:2} presents the properties of galaxy NGC 4254 as well as target new ULX, and this section also presents X-ray, optical, and infrared observations of new ULX. Details of these observations and data reduction and analysis are presented in Section \ref{sec:3}. Section \ref{sec:4} presents results and discussions of the properties of X-ray to IR of new ULX. Finally, Section \ref{sec:5} summarizes the main conclusions of this study.\\

\section{New ULX Candidate \& Observations}\label{sec:2}

\begin{figure*}
\begin{center}
\includegraphics[angle=0,scale=0.37]{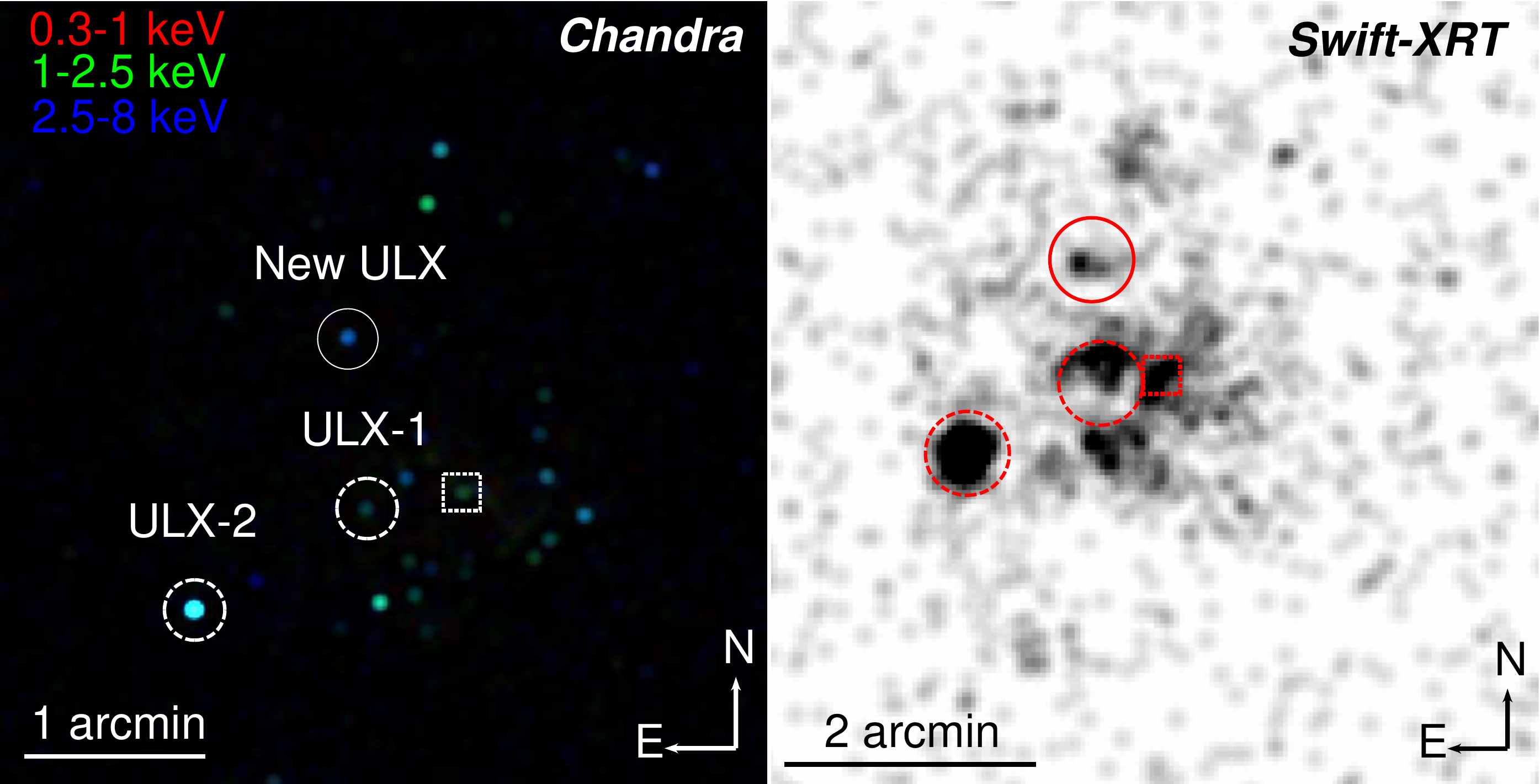}
\caption{The locations ULXs as well as new ULX (ULX-3) in NGC 4254 on the true color image of {\it Chandra} (left) and gray-scale stacked {\it Swift/XRT} (right) image. The energy ranges used for the true color image are highlighted on the {\it Chandra} image. The images are smoothed with a 3 arcsec Gaussian. In both panels, the dashed squares indicate the galaxy center.}
\label{location1}
\end{center}
\end{figure*}

\begin{figure*}
\begin{center}
\includegraphics[angle=0,scale=0.3]{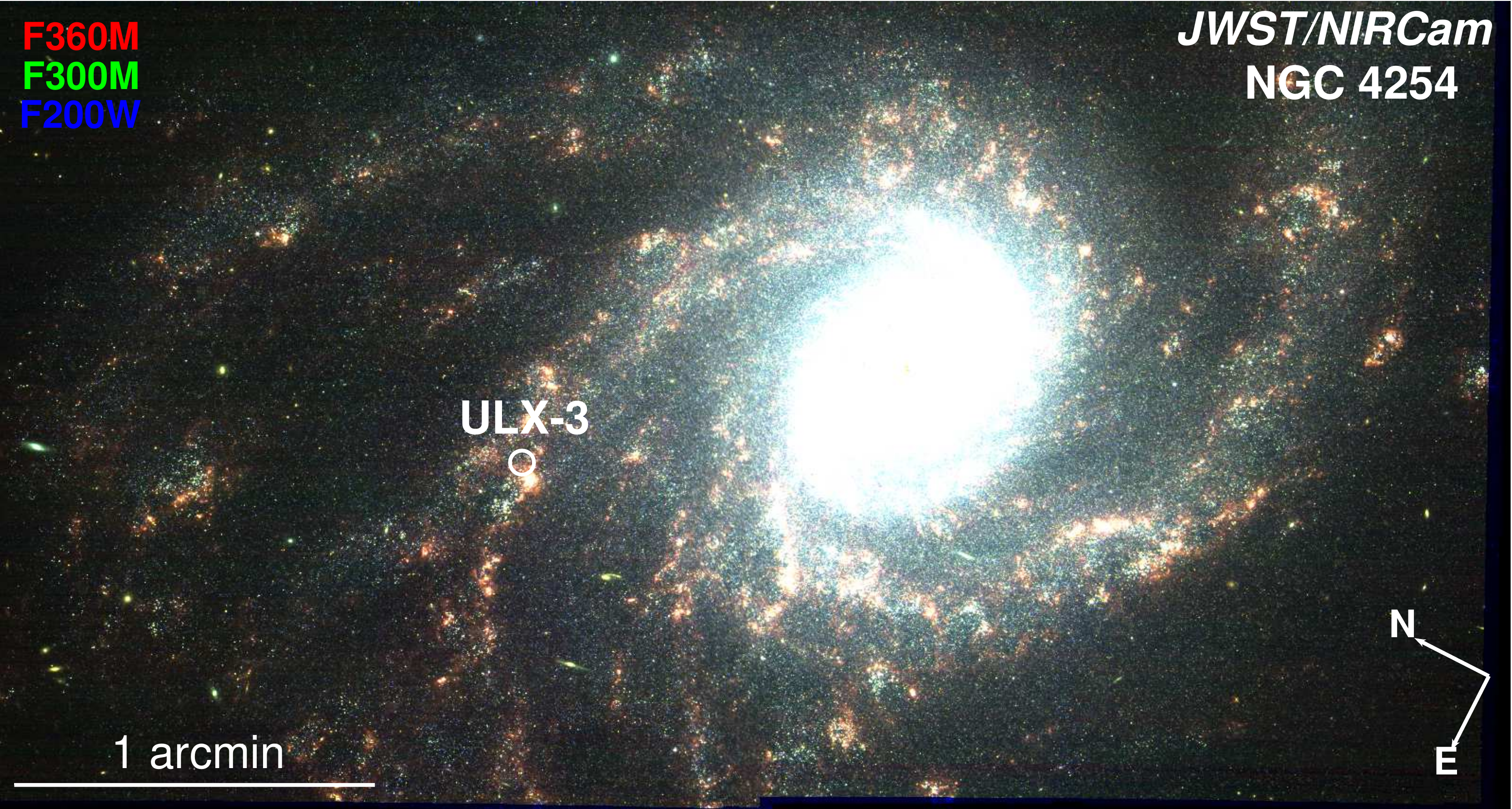}
\caption{The {\it JWST} RGB image of the galaxy NGC 4254. The filters used for RGB are highlighted on the image. The ULX-3 is indicated with a solid white circle.}
\label{location2}
\end{center}
\end{figure*}

Two ULX candidates have been reported by \cite{2005ApJS..157...59L} in nearby galaxy NGC 4254 (M99, VCC 0307) using the {\it ROSAT High-Resolution Imager}. Moreover, \cite{2022MNRAS.512.3284S} presented a highly luminous ULX source in this galaxy. However, there is no previously reported study for the X-ray source, which is the subject of this study. This source is not reported as ULX as the X-ray emission ({\it L$_{X}$}) did not exceed 10$^{39}$ erg s$^{-1}$ in previous {\it ROSAT}, {\it XMM-Newton} and {\it Chandra} observations. Therefore, this source will hereafter be called New ULX or ULX-3 as it has the third highest {\it L$_{X}$} radiation. The host galaxy NGC 4254 of the new ULX candidate is SA(s)c type face-on grand design spiral galaxy at a distance of 16.8 Mpc in the Virgo cluster \citep{2005ApJS..157...59L}. The location of this source is shown in Figures \ref{location1} and \ref{location2} on the {\it Chandra} and {\it Swift/XRT} and {\it JWST} images. As seen in Figure \ref{location2}, the new ULX is located in the spiral arms with the right accession (R.A) and declination (Decl.) of 184.719048 and 14.430346 in degree, respectively. Additionally, for the possibility that some of the brightest ULXs may be young supernova remnants (SNRs), for this SNRs catalogs (e.g. \citealp{2018ApJ...863..109Z} and \citealp{2022MNRAS.512.3284S}) were checked. Additionally, AGN (activated galactic nuclei) catalogs was also searched for NGC 4254 but, no SNRs or AGNs matching the new ULX candidate were found. \\

The galaxy NGC 4254 was observed on June 18, 2023, by using both the {\it JWST} near-infrared camera instrument ({\it NIRCam}) and mid-infrared instrument ({\it MIRI}) (GO program 2107, PI: J. Lee). The {\it JWST} observations include {\it NIRCam} imaging by using the F200W, F300M, F335M, and F360M filters and {\it JWST/MIRI} imaging using the F770W, F1000W, F1130W, and F2100W filters. NGC 4254 was also observed by {\it HST} on March 21, 2020, using WFC3/UVIS (The Wide Field Camera 3). The {\it HST} observations include UVIS imaging by using F275W, F336W, F438W, F555W, and F814W filters. Moreover, NGC 4254 was also observed three times during 45, and 5 ks ks by {\it Chandra} ACIS (Advanced CCD Imaging Spectrometer) in 2007, 2015, and 2021 and it was observed many times by {\it Swift/XRT} between 2010 and 2016. The details of observations are given in Table \ref{T:obs}.

\begin{table*}
\centering
\caption{The log of observations of the galaxy NGC 4254}
\begin{tabular}{ccccccccccccccccllll}
\hline
\\
Observatory/Instrument & Proposal ID & Date & Exp & Filter \\
& & (YYYY-MM-DD) & (s)\\
\hline
HST/WFC3/UVIS & 15654 & 2020-03-21 & 2250 & F275W \\
HST/WFC3/UVIS & 15654 & 2020-03-21 & 1950 & F336W \\
HST/WFC3/UVIS & 15654 & 2020-03-21 & 1062 & F438W \\
HST/WFC3/UVIS & 15654 & 2020-03-21 & 1055 & F555W \\
HST/WFC3/UVIS & 12118 & 2010-06-26 & 748 & F814W \\ 
\\
JWST/MIRI & 2107 & 2023-06-18 & 444 & F770W \\
JWST/MIRI & 2107 & 2023-06-18 & 610.5 & F1000W \\
JWST/MIRI & 2107 & 2023-06-18 & 1554.0 & F1130W \\
JWST/MIRI & 2107 & 2023-06-18 & 1609.5 & F2100W \\
JWST/{\it NIRCam} & 2107 & 2023-06-18 & 2405.0 & F200W \\
JWST/{\it NIRCam} & 2107 & 2023-06-18 & 773.1 & F300M \\
JWST/{\it NIRCam} & 2107 & 2023-06-18 & 773.1 & F335M \\
JWST/{\it NIRCam} & 2107 & 2023-06-18 & 858.9 & F360M \\
\hline
\\
& Target ID & Start time & N$^{a}$ & Exp$^{b}$ \\
& & (YYYY-MM-DD $-$ YYYY-MM-DD)& & (ks)& \\
\hline
Swift/XRT & 00031699 & 2010-04-20 $-$ 2010-04-22 & 2 & 5.2 \\
Swift/XRT & 00033132 & 2014-01-29 $-$ 2016-12-13 & 10 & 18.7 \\
Swift/XRT & 00091492 & 2013-02-21 $-$ 2014-02-10 & 6 & 3.6 \\
Swift/XRT & 00091993 & 2014-05-27 $-$ 2014-12-17 & 4 & 6.2 \\
\hline
\\
& Obs ID & Date & Exp \\
& & (YYYY-MM-DD) & (ks)& \\
\hline
Chandra/ACIS-S & 7863 & 2007-11-21 & 5.0 \\
Chandra/ACIS-I	& 17462 & 2015-02-16 & 45.0 \\
Chandra/ACIS-S & 24707 & 2021-04-14 & 5.0 &\\
\hline
Notes: $^{a}$ Number of observations $^{b}$ Total exposure time\\
\end{tabular}
\label{T:obs}
\end{table*}

\section{Data Reduction and Analysis} \label{sec:3}

\subsection{X-ray}

{\it Chandra} ACIS-S and ACIS-I observations were analyzed by using {\it Chandra Interactive Analysis of Observations} ({\scshape ciao})\footnote{https://cxc.cfa.harvard.edu/ciao/} 4.15.1
 software and calibration files {\scshape caldb} 4.10.4. The level 2 event files were obtained with {\it chandra\_repro}. For X-ray source detection, {\it wavdetect} task was used using longer exposure time observation (ObsID 17462). The coordinates of New ULX (hereafter ULX-3) were found as 12:18:52.59 and +14:25:49.12. ULX-3 was detected in two observations (ObsID 7863 and 17462). 
 
\begin{figure}
\begin{center}
\includegraphics[angle=0,scale=0.3]{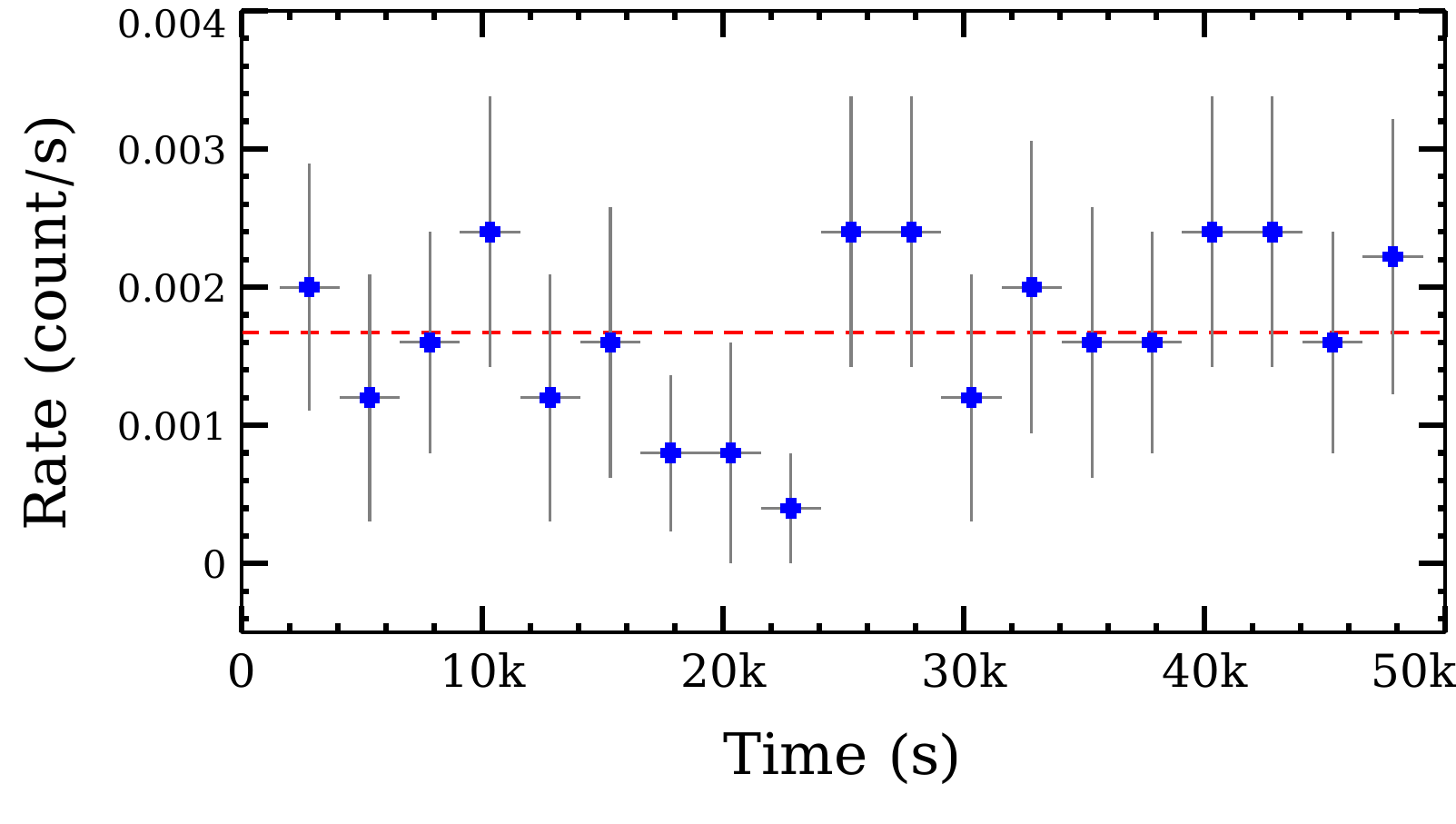}
\caption{The short-term light curve of ULX-3 using {\it Chandra} ObsID 17462 observation in 0.3-10 keV energy range. The dashed red line shows the mean value of count rates.}
\label{F:lcX}
\end{center}
\end{figure}

 The source and the background events were extracted from circular radii of 4 and 8 arcsec, respectively. The source spectra and light curves were obtained with the tasks {\it specextract} and {\it dmextract}, respectively. By using the task of {\it grppha} in {\it HEASOFT v 6.32.1} all spectra were grouped with 5 counts per energy bin due to low data quality, and the C-statistic was used for fitting to source spectra. Despite the presence of low-quality data,  single such as \textit{power-law} and \textit{multi-color disk blackbody} (\textit{diskbb}) and two-component models such as (\textit{power-law+diskbb}) were investigated with absorption components ({\it tbabs}) using {\scshape xspec} v12.13. Among the applied spectral models, an absorbed power-law model provided statistically acceptable fit to the only {\it Chandra} (ObsID17462) with a photon index ($\Gamma$) of 1.33 (see Table \ref{T:fit}). Moreover, the short-term count rate variability also was searched for ULX-3 using three {\it Chandra} data (ObsID 17462). For this, the light curves were binned intervals (e.g. 250, 500, 1000, and 2500 s) in the 0.3$-$10 keV energy band by using {\it axbary} and {\it dmextract}. The short light curve is displayed in Figure \ref{F:lcX} for long exposure time data while due to low data quality for the other two observations, light curves could not be created. \\

To source detection for {\it Swift/XRT} observations, the detection task of the {\it XIMAGE} in \textit{HEASOFT} tool was used with a signal-to-noise threshold of 3. The source ULX-3 was identified in only four of the {\it Swift/XRT} observations. Moreover, by using the online tool provided by \citep{2009MNRAS.397.1177E} the best source position was determined, and the light curve and spectrum of the source ULX-3 were generated. The 2SXPS catalog is used to identify sources in the field that should be excluded from the background. Therefore, the set of observations that are all covered by the 2SXPS catalog to use in the construction of light curves and spectra was selected. All products from the online tool are fully corrected for pile-up.\\

\begin{figure}
\begin{center}
\includegraphics[angle=0,scale=0.4]{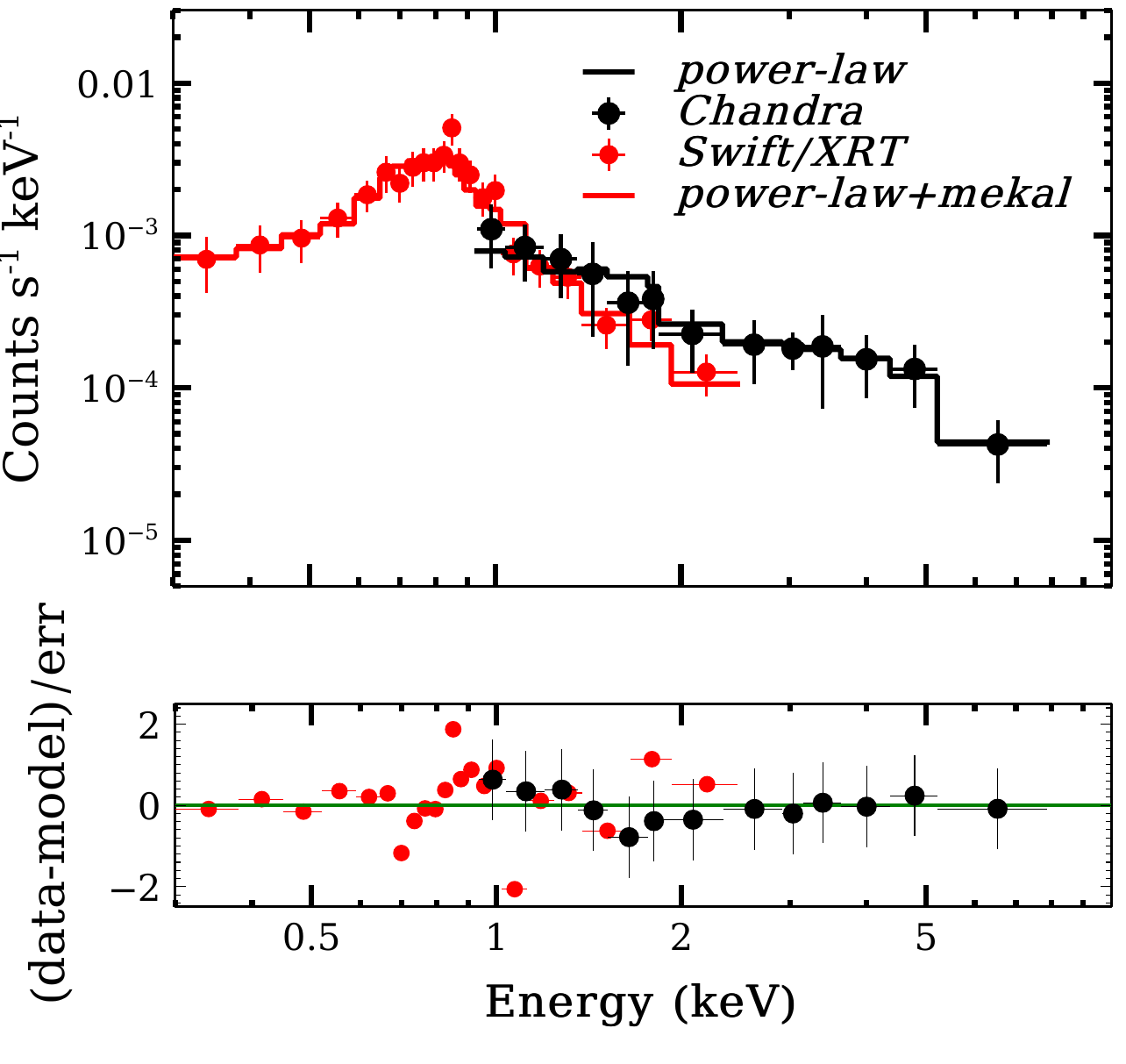}
\caption{The time-averaged \textit{Swift/XRT} spectrum of ULX-3 with two component model {\it power-law+mekal} (solid red line) and \textit{Chandra} (45 ks) spectrum of ULX-3 with one component model {\it power-law} (solid balck line).}
\label{F:model}
\end{center}
\end{figure}

By using the {\it grppha} all spectra were grouped with a minimum of 15 counts per energy bin. The C-statistic was used for fitting to source spectra. However, no acceptable model(s) could be obtained from individual observations due to short exposure times and poor data quality. On the other hand, the time-averaged spectrum was obtained using all available observations, but no statistically appropriate model for this spectrum was obtained in the energy range of 0.3-10 keV. However, especially in the energy range 0.3-2.2 keV, the spectrum well above the background level was adequately well-fitted by the two-component {\it power-law+mekal} model with reduced chi-square $\chi^2_{\nu}$ of 0.80. According to this model, the {\it mekal} temperature was found to be 0.40 keV and the {\it power-law (po)} photon index 1.53 with one absorption component of 0.01 $\times$ 10$^{22}$ cm$^{-2}$. On the other hand, even though a number of models consisting of \textit{diskbb}, \textit{blackbody} or combinations of them were tried to be fitted, acceptable fitting statistics and/or physically meaningful parameters could not be obtained. Moreover, The time-averaged spectrum and the \textit{Chandra} energy spectrum were also fitted together including a constant parameter, but no acceptable fit statistics were obtained. The time-averaged and \textit{Chandra} energy spectrum is displayed in Figure \ref{F:model}. All the spectral fitting results for both \textit{Chandra} and \textit{Swift/XRT} are summarised in Table \ref{T:fit}.

\begin{table*}
\centering
\begin{minipage}[b]{0.9\linewidth}
\caption{Spectral parameters obtained with one and two-component model fits for ULX-3}
\begin{tabular}{cccccccccccccc}
\hline
&& \multicolumn{3}{|c| }{Time-average spectrum of \textit{Swift/XRT}} &&&&& \multicolumn{4}{|c|}{\textit{Chandra} (ObsID17462)} \\
Parameter & unit & \textit{diskbb+po} & \textit{bb+po} & \textit{mekal+po} &&&&&& \textit{po} & \textit{diskbb} & \textit{mekal+po}\\
\hline
$N_{H}$ & $10^{22}$ cm$^{-2}$ & 0.11 & 0.51 & $0.01_{-0.01}^{+0.01}$ &&&&&& <0.01 & <0.01 & $1.10_{-1.15}^{+1.67}$\\
kT & keV & ... & 35 & $0.40_{-0.10}^{+0.13}$ &&&&&& ... & ...& $0.11_{-0.11}^{+2.06}$ \\
Tin & keV & 0.22 &... &... &&&&&&... &$2.56_{-1.05}^{+28.67}$ & ...\\
$\Gamma$ &  & 2.90 & 6.87 & $1.53_{-0.30}^{+0.38}$ &&&&&& $1.33_{-0.18}^{+0.21}$ & ... & $1.43_{-0.94}^{+1.11}$ \\
$L_{\mathrm{X}}$ & $10^{39}$ erg s$^{-1}$ & 2.69 & 2.65 & $2.76_{-0.09}^{+0.11}$ &&&&&& $0.88_{-0.04}^{+0.08}$ & $0.81_{-0.12}^{+0.16}$ & $0.71_{-0.21}^{+0.33}$\\
C-stat/dof & & >2 & >2 & 0.80 &&&&&& 0.64 & 0.75 & 0.40\\
\hline
\end{tabular}
\\ Notes: Luminosity values were calculated at 0.3$-$10 keV energy range and  all errors are at the confidence range of 2.706. \textit{po=power-law}; \textit{bb=blackbody}; \textit{diskbb=multi-color disk blackbody}.\\
\label{T:fit}
\end{minipage}
\end{table*}

\begin{figure*}
\begin{center}
\includegraphics[angle=0,scale=0.35]{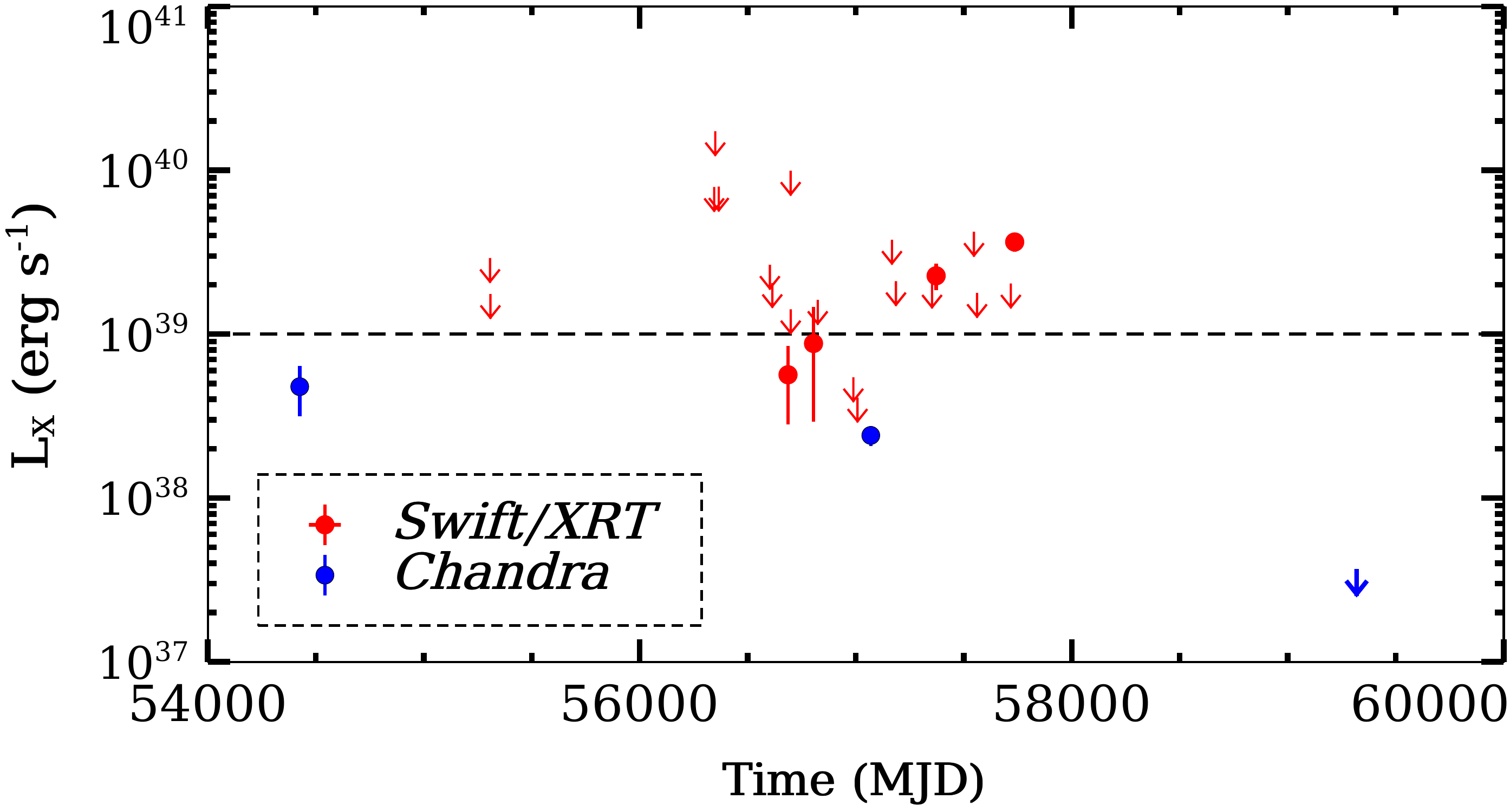}
\caption{The 0.3–10 keV X-ray long-term light-curve of ULX-3 obtained from {\it Chandra} (filled blue circles) and {\it Swift/XRT} (filled red circles) observations. 3-$\sigma$ upper limits for {\it Chandra} and {\it Swift/XRT} data are shown by the blue and red downward arrows. The dashed line represents ULX-state.}
\label{F:LC}
\end{center}
\end{figure*}

\begin{figure}
\begin{center}
\includegraphics[width=\columnwidth]{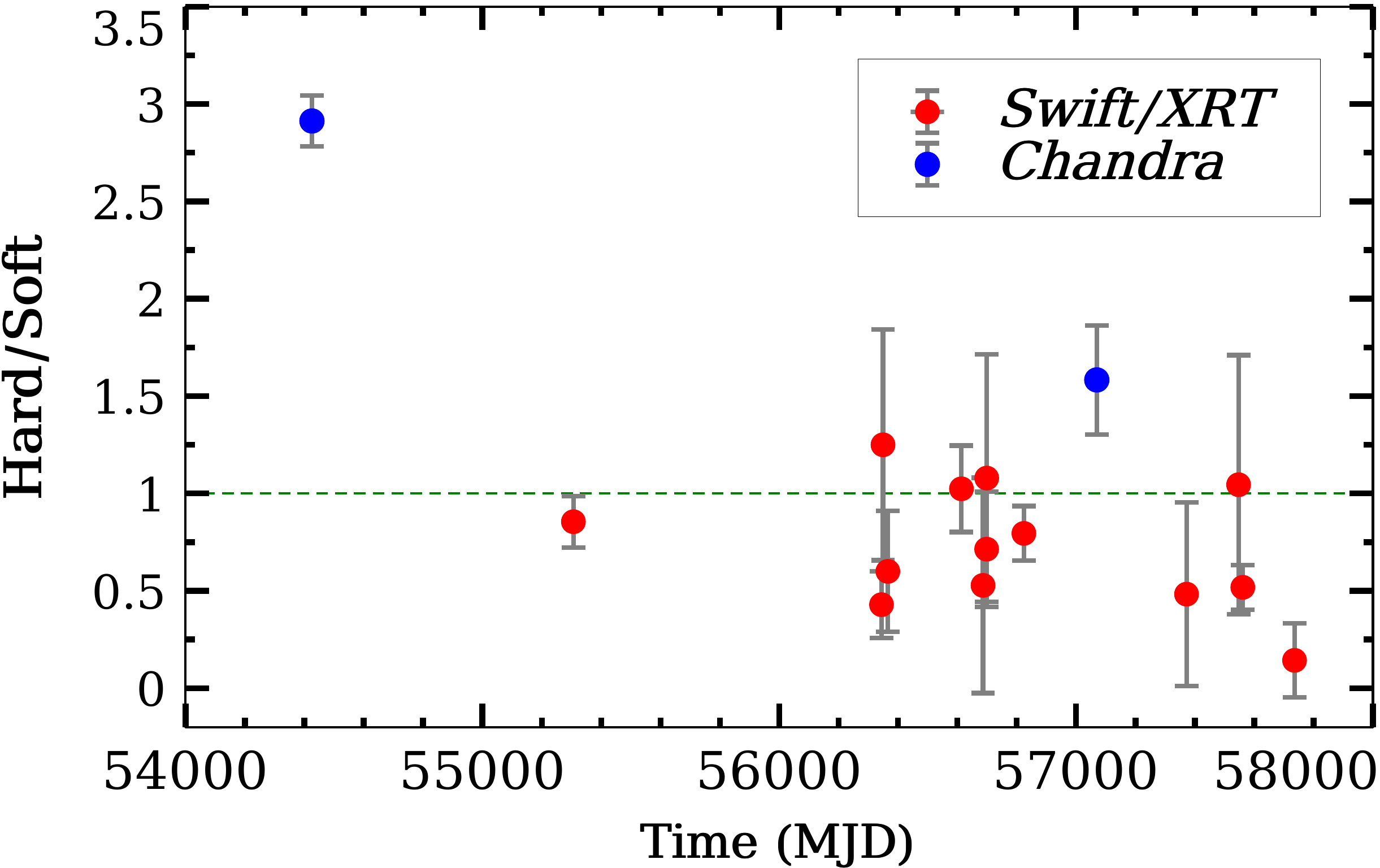}
\caption{Variations in the spectral hardness ratios of ULX-3 with time. The hardness ratios (Hard/Soft) were derived in two energy bands, Soft = 0.3-1.5 keV and Hard = 1.5-10 keV. {\it Chandra} and {\it Swift/XRT} data are shown by blue and red filled circles.}
\label{F:hardness}
\end{center}
\end{figure}

In order to obtain the luminosities of ULX-3, the unabsorbed X-ray fluxes were derived using all available X-ray observations in the energy range of 0.3-10 keV based on the distance of NGC 4254 is 16.8 Mpc. For this, count rates including 3-$\sigma$ upper limits, where ULX-3 was not detected in {\it Swift/XRT} observations were converted to unabsorbed fluxes using a spectral shape obtained best fitting model parameters of the time-averaged spectrum, {\it mekal} temperature of  \textit{kT}=0.4 keV and absorption component, N$_{H}$ =0.01 $\times$ 10$^{22}$ cm$^{-2}$ with WebPIMMS\footnote{https://heasarc.gsfc.nasa.gov/cgi-bin/Tools/w3pimms/w3pimms.pl}. The source ULX-3 was detected in only four of the 22 {\it Swift/XRT} observations and the range of X-ray luminosity was derived as $\sim$ 5 $\times$ 10$^{38}$ $-$ 4 $\times$ 10$^{39}$ erg s$^{-1}$. In the case of 3-$\sigma$ upper limits, luminosity range of ULX-3 was derived as $\sim$ 2 $\times$ 10$^{38}$ $-$ 10$^{40}$ erg s$^{-1}$. The ULX-3 was detected in two out of three {\it Chandra} observations while in the remaining observation was detected as 3-$\sigma$ upper limit. The luminosity range in two observations was derived as $\sim$ (1 $-$ 4) $\times$ 10$^{38}$ erg s$^{-1}$. For the 3-$\sigma$ upper limit, the count rate was converted to unabsorbed flux using \textit{power-law} model parameter of $\Gamma$ = 1.33  and  N$_{H}$ =0.01 $\times$ 10$^{22}$ cm$^{-2}$ obtained from the \textit{Chandra} energy spectrum with \textit{srcflux} in {\scshape ciao}. Unabsorbed X-ray flux for 3-$\sigma$ upper limit was derived as $\sim$ 2 $\times$ 10$^{37}$ erg s$^{-1}$. The long-term X-ray luminosity light curve of the ULX-3 is plotted in Figure \ref{F:LC}. To examine the energy distribution of the X-ray emission from ULX-3, the hardness ratios (H/S) were plotted based on two energy bands: a soft band (S): 0.3–1.5 keV, and a hard band (H): 1.5–10 keV (see Figure \ref{F:hardness}). \\

\subsection{Optical \& Infrared}

\subsubsection{Source Detection and Photometry}

{\scshape iraf} ({\it Image Reduction and Analysis Facility})\footnote{https://iraf-community.github.io/} was used throughout this study for source detection and photometry. Prior to source detection, {\it imstat} and {\it imexamine} tasks were used to estimate the background for each filter. For all images, point-like sources were detected at 3-$\sigma$ detection threshold with the {\it daofind} task, and aperture photometry was performed using the {\it DAOPHOT} package. To perform aperture photometry a circular aperture with a radius of 3 pixels was chosen. The local background level around each source was estimated using a circular annulus with an inner radius of 7 pixels and an outer radius of 12 pixels. The Vega magnitudes for {\it HST} images were derived from instrumental magnitudes using {\it WFC3} zero point magnitudes from \cite{2022AJ....164...32C}. In the case of {\it JWST} images, Vega magnitudes (hereafter Vega$_{mag}$) were obtained from flux densities in MJy/sr using the following equations\footnote{https://jwst-docs.stsci.edu/jwst-near-infrared-camera/nircam-performance/nircam-absolute-flux-calibration-and-zeropoints}:

\begin{equation}
Vega_{mag} = -2.5 \times log(f1 \times PIXAR\_SR/f2)
\end{equation}
where PIXAR\_SR is the average area of a pixel in sr (steradian) obtained from FITS header keyword PIXAR\_SR and {\it f1} and {\it f2} are flux densities in MJy/sr and Vega$_{flux}$ in MJy, respectively. {\it NIRCam} images in units of MJy/sr are derived from data rate images in units of counts/s (DN/s). The conversion factor is obtained from the FITS header under the PHOTMJSR, keyword, which has units of (MJy/sr) / (DN/s). Therefore,

\begin{equation}
flux(DN/s) = flux[MJy/sr] / PHOTMJSR.
\end{equation}

\begin{equation}
instrumental_{mag} = -2.5 \times log(flux[DN/s])
\end{equation}

\begin{equation}
Vega_{mag} = instrumental_{mag} + ZP_{Vega}
\end{equation}

\begin{equation}
ZP_{Vega} = -2.5 \times log10(PHOTMJSR \times PIXAR\_{SR}/f2)
\end{equation}

Finally, to derived ZP$_{Vega}$ (zeropoints Vega magnitudes), Vega$_{flux}$ ({\it f2}) is obtained from {\it Spanish Virtual Observatory} filter profile service\footnote{http://svo2.cab.inta-csic.es/theory/fps/}. To derive the aperture corrections with a radius between 3 pixels to 10 pixels for each image using 30 isolated and bright sources for both observatories. The obtained Vega magnitude values were corrected with the extinction E(B-V) = 0.67 mag from \cite{2018ApJ...863..109Z}. Derived aperture correction values, and also ZP$_{Vega}$ are given in Table \ref{T:fotometry}.

\subsubsection{Identification of Optical and IR Counterpart of ULX-3}

Relative astrometry was performed between the \textit{Chandra}, \textit{JWST}, and {\it HST} images to identify the possible optical and IR counterparts of ULX-3.
Since the \textit{GAIA} source catalog contains many astrometrically corrected objects such as AGNs, it is a good tool for determining the multi-wavelength counterparts of ULXs. For these, references for astrometric calculations were searched by comparing \textit{GAIA/DR3}\footnote{https://www.cosmos.esa.int/web/gaia/dr3} sources with \textit{Chandra} X-ray sources. Two possible references were identified, one of which is the galaxy nuclei and the other is the ULX-1 detected with a low signal-to-noise ratio in the \textit{Chandra} image. However, these sources were not used for astrometric calculations since the possible references have different shift directions. In the case of the \textit{Two Micron All Sky Survey (2MASS)} source catalog, two sources, the galaxy nuclei and the X-ray source identified in previous studies as ULX-1, were considered as possible references. However, because of the different directions of the shifts, they are not convenient references for astrometric calculations.

\begin{figure}
\begin{center}
\includegraphics[width=\columnwidth]{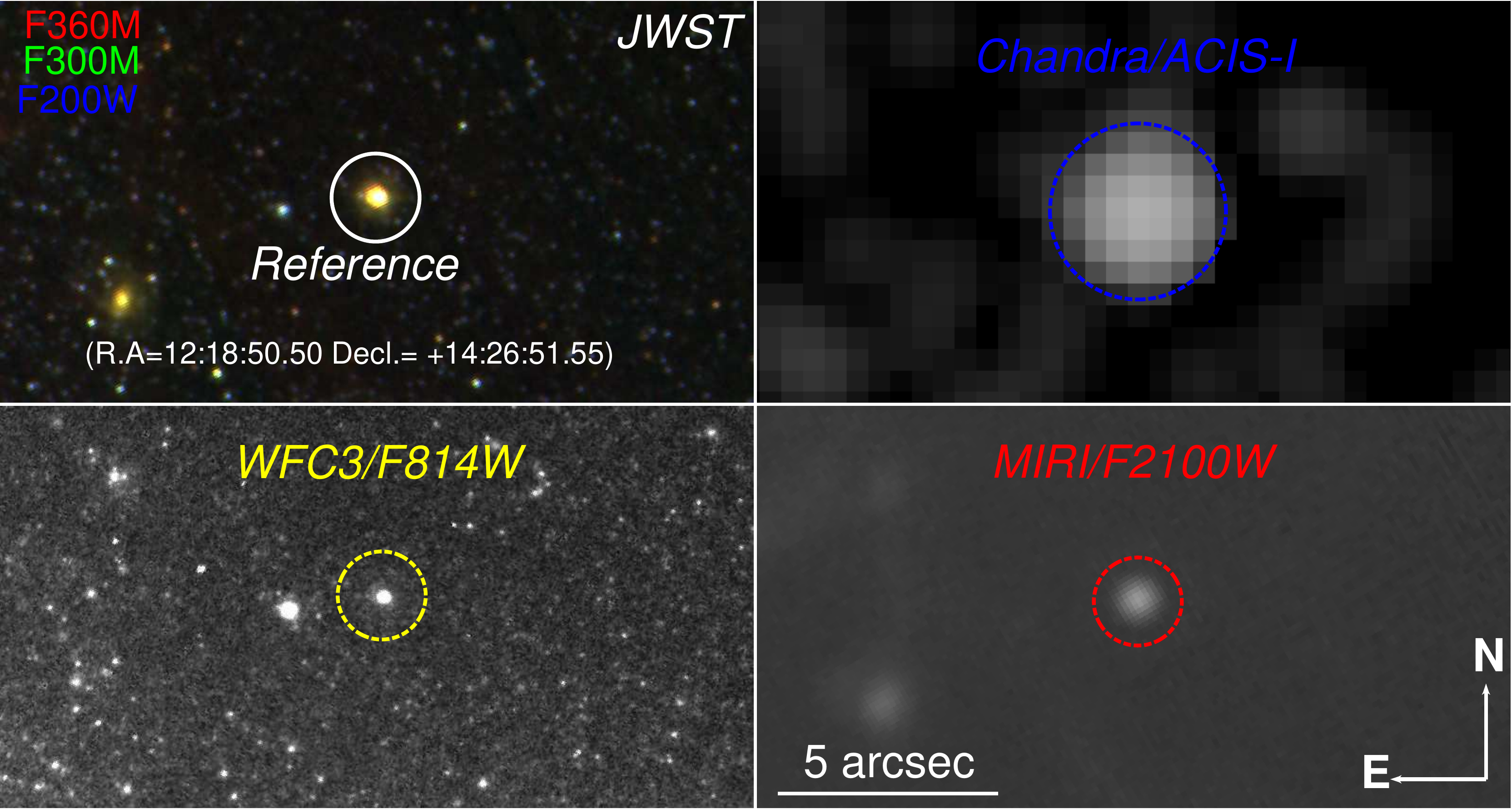}
\caption{The location of reference source used for astrometric calculations shows on the {\it NIRCam} RGB, {\it Chandra}, {\it HST/WFC3} F814W, and {\it JWST/MIRI} F2100W images from left to right, respectively.}
\label{F:ref}
\end{center}
\end{figure}

Since there were no acceptable references from similar catalogs, \textit{Chandra} ($\sim$ 0.5 arcsec/pixel) and \textit{JWST/NIRCam} F200W ($\sim$ 0.03 arcsec/pixel) images with good spatial resolution were used to identify counterparts of ULX-3. AGN candidate and the galactic nucleus, which have the same shift directions, two were selected as two references between \textit{Chandra} and \textit{JWST/NIRCam} images. The AGN candidate was detected with a high signal-to-noise ratio in images taken by all instruments (see Figure \ref{F:ref}). Between {\it Chandra} and the NIRcam/F200W images, shifts of 0.01 arcsec and 0.16 arcsec for R.A and Decl were found. The total astrometric error between {\it Chandra} and {\it JWST/NIRCam} was derived as 0.17 arcsec with 1-$\sigma$ significance. Finally, accepted 1-$\sigma$ accuracy in {\it Chandra} position as 0.1 arcsec, the positional error radius of ULX-3 was derived as 0.55 at 95\% confidence level. Within this error radius point-like a unique IR source was found. The corrected X-ray position of the IR counterpart was found to be R.A = 12:18:52.585 and Decl = +14:25:49.183 for the F200W image. In addition, to check whether there is also a shift between the \textit{NIRCam} and the \textit{MIRI} images, relative astrometry was performed between the images of the two instruments. Shifts of 0.07 arcsec -0.02 arcsec were found between the two instrument images for R.A and Decl., respectively. After aligning the images, the NIR counterpart was also detected in \textit{JWST/MIRI} images. \\

\begin{figure*}
\begin{center}
\includegraphics[angle=0,scale=0.3]{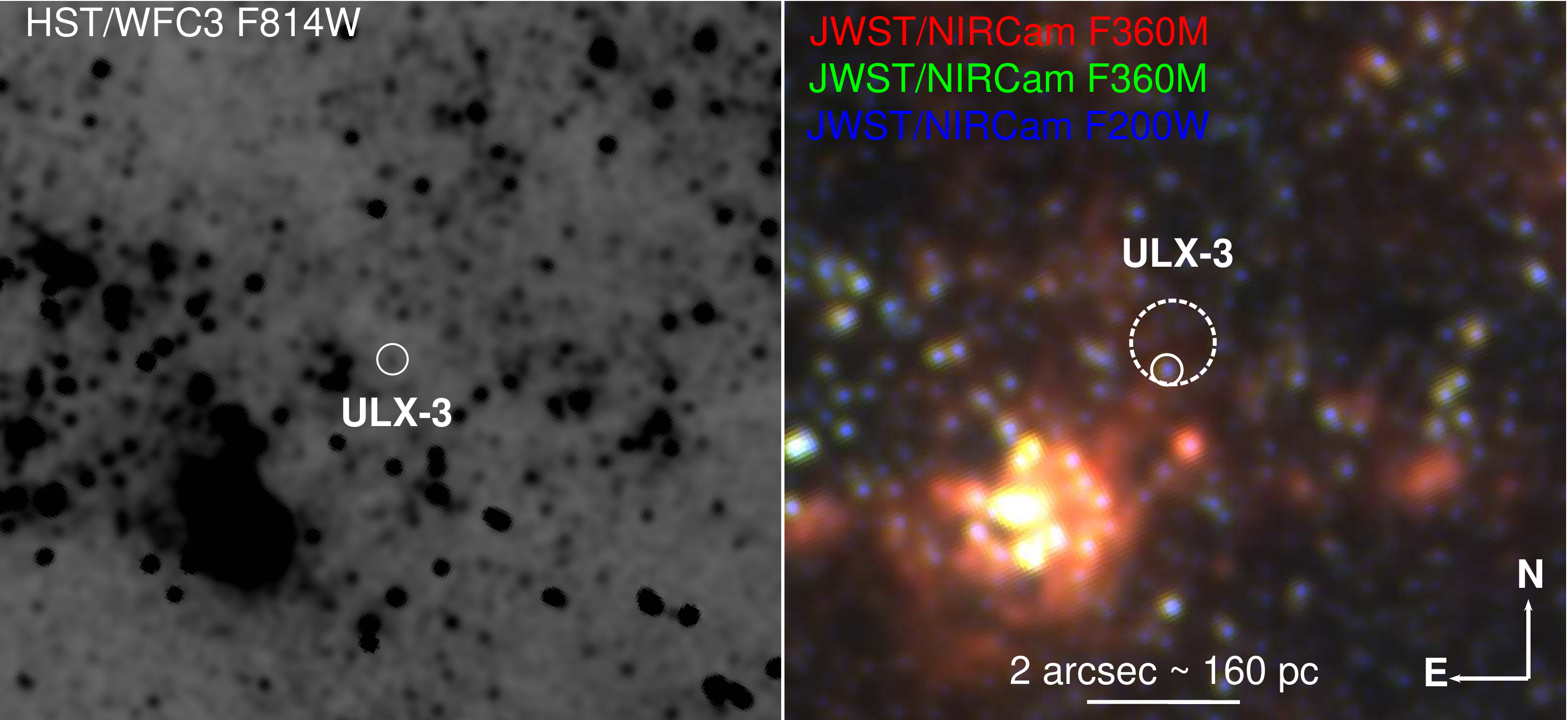}
\caption{The location of counterpart ULX-3 on the {\it HST/WFC3} F814W (left) image and RGB (Red: Green: Blue) {\it JWST/NIRCam} image (right). The filters used for RGB image are highlighted on the image. Dashed and solid white circles represent the astrometric error radius and counterpart, respectively. The images are smoothed with 3 arcsec Gaussian. Both images are on the same scale.}
\label{F:pos}
\end{center}
\end{figure*}

Moreover, relative astrometry was performed between {\it NIRCam} and {\it HST} images to determine the optical counterpart(s). The astrometric offsets between {\it JWST/NIRCam} and {\it HST/WFC3} were derived as -0.11 arcsec and -0.12 arcsec at 1-$\sigma$ significance level for R.A and Decl., respectively. The astrometric error radius for this calculation was derived as 0.34 arcsec at 95\%. In addition, shifts between each {\it WFC3} filter were checked, and source positions were corrected according to shift values. Within this error radius, a unique optical counterpart was determined only in the F814W filter with R.A and Decl. of 12:18:52.560 and +14:25:49.083 while for other filters this counterpart was not detected. The location of this optical counterpart coincides with the position of the IR counterpart mentioned above. The corrected X-ray positions of the optical and IR counterparts (hereafter counterpart) of ULX-3 are shown in Figure \ref{F:pos}. The corrected Vega$_{mag}$ of counterpart are given in Table \ref{T:fotometry}.\\

\begin{table}
 \caption{The zero point magnitude (ZPM), aperture corrections, and dereddened Vega magnitudes of the counterpart ULX-3}
 \begin{tabular}{cccccll}
\hline\hline
Instruments & Filter & ZPM & Correction & Magnitude \\
 &&(Vega) & (Vega) & (Vega) \\
\hline
WFC3 & F814W & 25.61 & 0.28 & 25.74 $\pm$ 0.27\\
NIRCam & F200W & 25.56 & 0.32 & 20.60 $\pm$ 0.01 \\
NIRCam & F300M & 23.97 & 0.34 & 20.04 $\pm$ 0.03 \\
NIRCam & F335M & 23.78 & 0.33 & 19.91 $\pm$ 0.02 \\
NIRCam & F360M & 23.69 & 0.36 & 19.31 $\pm$	0.02 \\
MIRI & F770W & 22.19 & 0.42 & 16.98 $\pm$ 0.02 \\
MIRI & F1000W & 21.41 & 0.44 & 15.05 $\pm$ 0.02 \\
MIRI & F1300W & 19.86 & 0.51 & 14.47 $\pm$ 0.03 \\
MIRI & F2100W & 19.68 & 0.54 & 13.18 $\pm$ 0.01 \\
\hline
\label{T:fotometry}
\end{tabular}
\end{table}

\subsubsection{CMD and SED}\label{secsed}

\begin{figure*}
\begin{center}
\includegraphics[angle=0,scale=0.4]{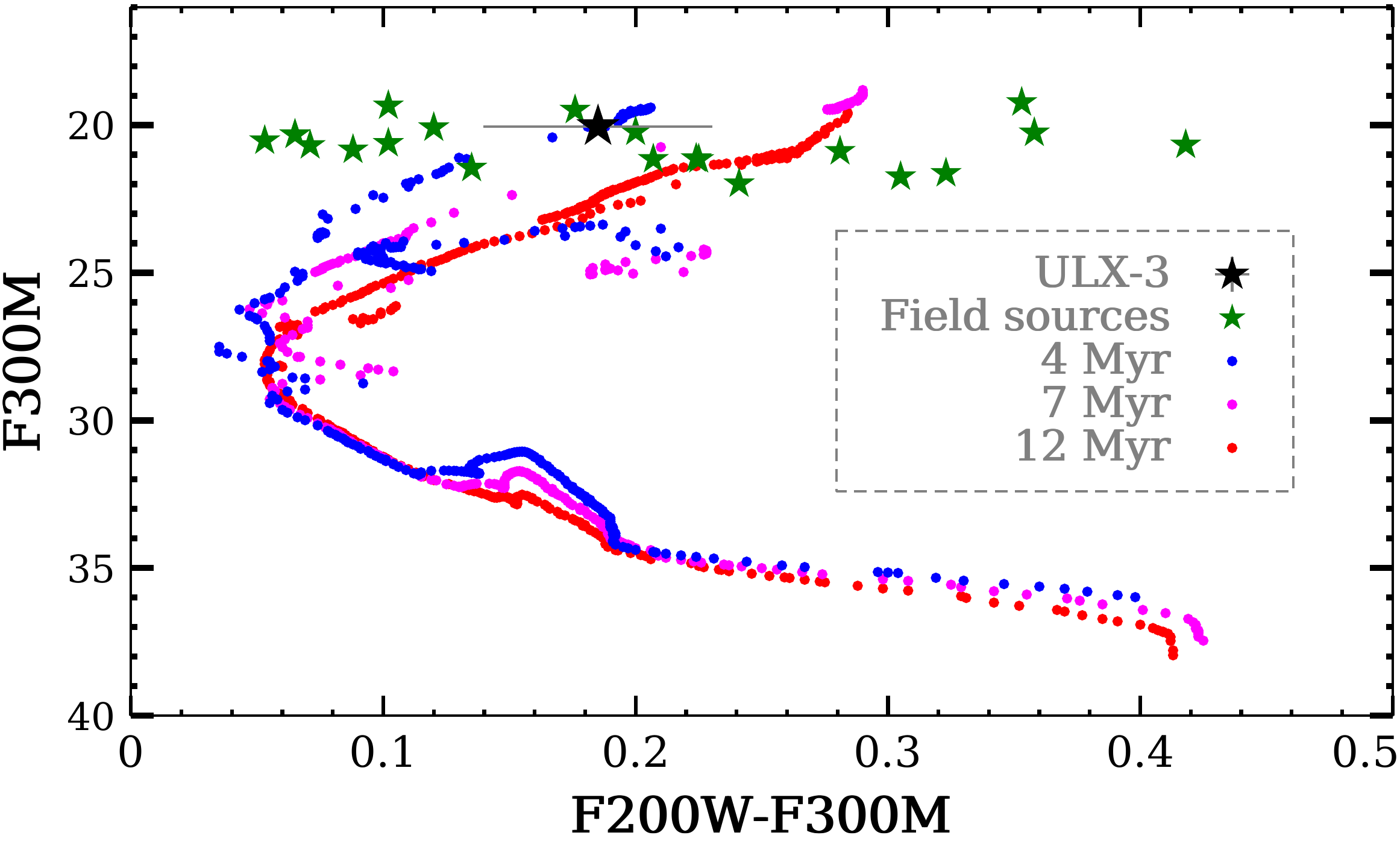}
\includegraphics[angle=0,scale=0.45]{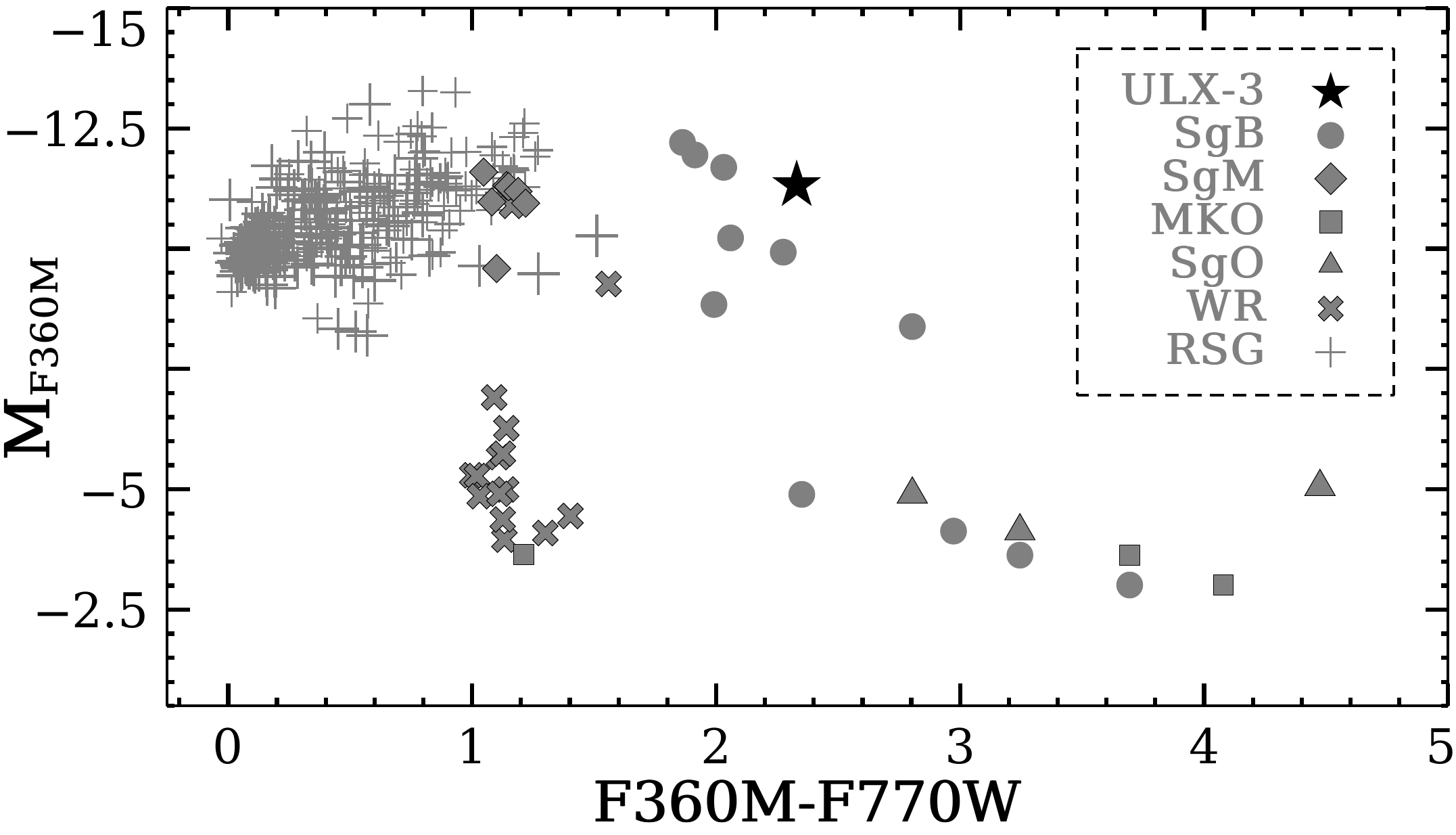}
\caption{Top: {\it JWST/NIRCam} CMD for field stars as well as the counterpart of ULX-3. Field star candidates and counterpart ULX-3 are represented with green and black stars, respectively. Bottom: {\it JWST} color-magnitude diagram of counterpart ULX-3 (filled red circle) plotted over massive stars and RSGs in the LMC and RSGs in SMC of known spectral type compiled by \citealp{2009AJ....138.1003B}, \citealp{2011ApJ...727...53Y} and \citealp{2012ApJ...754...35Y}. All source labels are plotted on the figure. ({\it SgB:supergiant B(e); SgM: supergiant M-type; MO: main sequence O-stars; SgO:supergiant O stars; WR: Wolf-Rayet stars; RSG: Red supergiant stars})}
\label{F:CMD}
\end{center}
\end{figure*}

To estimate the possible age of the counterpart many point-like NIR sources around ULX-3 were examined by using the {\it JWST/NIRCam} images (see the left panel of Figure \ref{F:CMD}). A color-magnitude diagram (CMD) in F300M versus F200W-F300M was produced. This CMD over-plotted with Vega mag PARSEC \cite{2012MNRAS.427..127B}
isochrone. The PARSEC isochrones were corrected for extinction E(B-V) of 0.67 mag, distance modulus of 31.12 mag, and solar metallicity (Z) of 0.02. The CMD of the counterpart with 19 NIR sources represented by corrected isochrones is shown in the top panel of Figure \ref{F:CMD}. It is apparent that field star candidates are within the age range of 4-12 Myr. The mass range corresponding to these ages is 18-30 M${\odot}$. If the donor star is dominant for the observed NIR radiation and also compared to the field stars, the age and mass of the counterpart should be 5 Myr and 28 M${\odot}$, respectively.\\

By using the catalog of IR massive stars and RSGs properties of Large Magellanic Cloud (LMC) and Magellanic Cloud (SMC) compiled by \cite{2009AJ....138.1003B}, \cite{2011ApJ...727...53Y}, and \cite{2012ApJ...754...35Y}, respectively we can achieve important constraints on possible donor stars of ULXs from their IR color-magnitude diagrams \citep{2019ApJ...878...71L} and therefore, another CMD using these catalog was produced. The photometric results of the 1750 sources for massive stars and 191 RSG candidates in the catalog were obtained from {\it Spitzer} mid-IR images with a wavelength range of 0.3 to 24 $\mu$m (3000-240000 \AA). If the counterpart ULX-3 of color magnitude is plotted over the CMD of these sources, we can constrain the spectral type of the possible donor. In terms of wavelengths and similar colors in the catalog, wavelengths 3.6 $\mu$m and 8.0 $\mu$m consistent with ULX-3 were chosen, which partially correspond to wavelengths {\it NIRCam} 3.6 $\mu$m (F360M) and {\it MIRI} 7.7 $\mu$m (F770W). As seen in the bottom panel of Figure \ref{F:CMD}, the color of ULX-3 is close to the population of bright SgB (supergiant B(e) stars) ([3.6$\mu$m]-[7.7$\mu$m] = 1.9-2.4 mag). \\ 

\begin{figure*}
\begin{center}
\includegraphics[angle=0,scale=0.4]{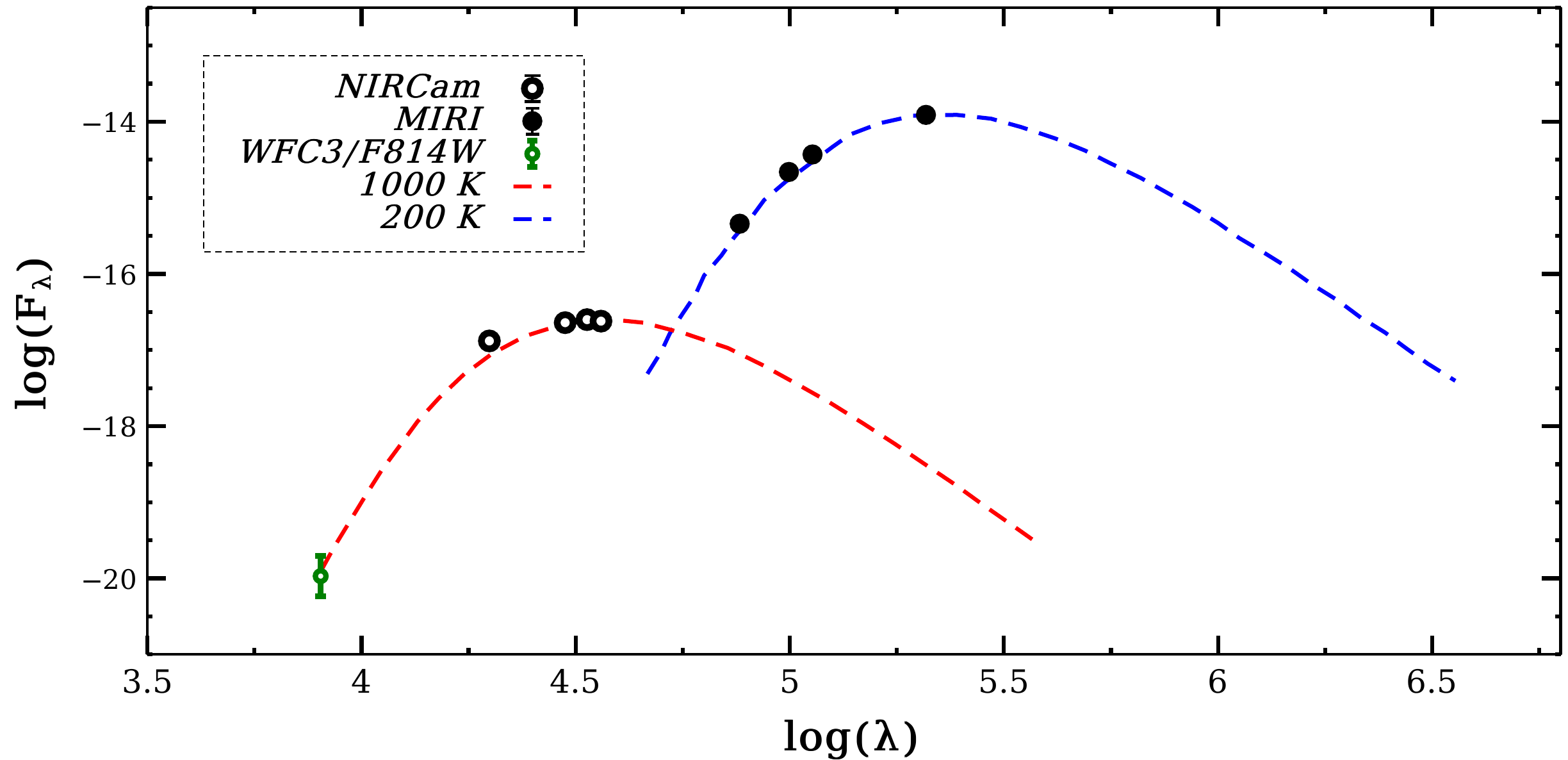}
\caption{SED constructed in Section \ref{secsed}. The {\it HST/WFC3} F814W, {\it JWST/NIRCam} data are represented by unfilled green and black circles while {\it JWST/MIRI} data are represented by filled black circles, respectively. The dashed red and blue lines show the blackbody models with temperatures of 1000 K and 200 K. The units of y and x axes are erg s$^{-1}$ cm$^{-2}$ and \AA, respectively.}
\label{F:SED}
\end{center}
\end{figure*}

To obtain the SED (spectral energy distribution) of counterpart ULX-3 the flux values are derived from Table \ref{T:fotometry}. Wavelengths are considered pivot values for each filter. As can be clearly seen in Figure \ref{F:SED}, the SED of the counterpart is well represented by two different blackbody models for NIR and mid-IR emission. The NIR emission of counterpart ULX-3 was well-fitted by the {\it blackbody} {\it T} $\sim$ 1000 K while mid-IR emission was well-fitted by {\it blackbody} $\sim$ 200 K. The reduced chi-square, $\chi^2_{\nu}$, value of the model is 0.89 with the number of degrees of freedom (dof) of 5.

\section{Results and Discussions}\label{sec:4}

The new transient ULX-3 candidate was analyzed in detail from the mid-IR to the X-ray band. By performing precise astrometry, a unique optical and IR counterpart of ULX-3 was identified. In this section, X-ray and optical-IR results are discussed under two headings.

\subsection{X-ray}

ULX-3 has been seen as a potential transient XRB candidate source with X-ray luminosity around 10$^{38}$ erg s$^{-1}$ in two of the three {\it Chandra} observations. ULX-3 has been observed 22 times between 2010 and 2016 by {\it Swift/XRT}, but it was detected in only four of these 22 observations. The X-ray luminosity of ULX-3 exceeded 10$^{39}$ erg s$^{-1}$. Furthermore, all available catalogs for the galaxy NGC 4254 were searched and no source (such as AGN and SNR) matching ULX-3 was found, therefore it was identified as a transient ULX candidate. Taking into account all of the X-ray observations, ULX-3 varies by more than two orders of magnitude in the energy range of 0.3-10 keV. This high variability is one of the typical properties of transient ULX candidates \citep{2018MNRAS.476.4272E,2018ApJ...864...64H,2019MNRAS.486.5709L,2022MNRAS.510.4355A}, but no evidence was found for short-term (e.g. for each exposure time) count rate variability for ULX-3. Some HMXBs with Be-star donors can occasionally produce very high mass transfer rates in the short term due to dynamic effects on the Be-star disk and these systems are very good examples of repeated transitions from a normal XRB to ULX and ULX to normal XRB \citep{2023NewAR..9601672K}.\\

As can be seen in the hardness ratios diagram plotted in Figure \ref{F:hardness}, transitions for ULX-3 were observed from hard to soft and then from soft to hard over time (very hard to soft to hard to soft), but it is seen that it is mostly in a soft state. These transitions are typical of Galactic HMXBs \citep{2017ARA&A..55..303K}. Moreover, as seen in the time-averaged spectrum, the source is soft at high energies where soft X-rays dominate. In this case, the soft emission may originate from the photosphere of the optically thick wind \citep{2021ApJ...906...36Q} or spectral transitions may be
caused by Lense–Thirring \citep{2018MNRAS.475..154M}. Such transitions have been observed in neutron stars, especially for atoll sources and black hole LMXBs and also ULXs in different shapes in hardness-intensity diagrams (HID) \citep{2005ApJ...633..358A,2007A&ARv..15....1D,2021A&A...651A..54M,2023arXiv230800141A}. The specific q-shape track \citep{2005A&A...440..207B,2022MNRAS.511.2535S} has been observed in HID of black hole XRBs and for NS with island and banana shape track \citep{2005ApJ...633..358A} but, q-shape and/or island and banana states for ULX-3 was not found. On the other hand, It is possible that ULX-3 may show such transitions in the HID since we have a limited number of observations obtained thus far. Finally, a scenario transitions for ULX-3, during a period when the accretion rate increases, more mass falls into the accretion disk, which may lead to more soft X-ray emission. In other words, periods when the mass transfer and the geometry of the accretion disk change can be observed as transitions.\\

Due to poor data quality or lack of data, it is not very clear which model represents the energy spectra of ULX-3. However, the time-averaged spectrum obtained using 22 {\it Swift/XRT} observations is adequately represented by the two-component {\it power-law+mekal} model with {\it mekal} temperature of 0.4 keV and $\Gamma$ of 1.53. In the two-component energy spectrum of ULX-3, 74 percent of the observed radiation comes from the {\it mekal} component. The {\it mekal} model is the best proof of thermal plasma and hot diffuse gas \citep{1996uxsa.conf..411K}. In the literature, the {\it mekal} component has been reported for many X-ray sources: e.g; ULXs \citep{2005ApJ...630..228K,2005ApJ...635..198D,2013MNRAS.435.1758S,2016MNRAS.456.3840E,2016ApJ...831...56U,2022MNRAS.510.4355A}; Be stars \citep{2007A&A...474..983L,2013ApJ...765...13T}; AGNs \citep{2010ApJ...711..144N,2020A&A...638A..67F}. The spectrum of the local background also was extracted which may indicate that the {\it mekal} component is not due to the background. Unlike \textit{Swift/XRT}, ULX-3 has a low/high state in the \textit{Chandra} energy spectrum, as shown in Figure \ref{F:model}, and its spectrum is represented by the \textit{power-law} model. The \textit{Chandra} data do not have much sensitivity below 1 keV due to the contamination of the detectors, which may be why the \textit{Chandra} energy spectrum only requires a \textit{power-law}. In addition, the {\it mekal} component may represent emission from the hot collisionally ionized gas in the region surrounding ULX-3. In the case of {\it power-law} component may represent non-thermal emission from the coronal region.

\subsection{Optical-IR}

Through precise astrometric calculations, a unique optical and IR counterpart for ULX-3 was identified. This counterpart is detected only in the F814W ($\sim$ R-band) filter at optical wavelengths, while it is detected in all NIR and mid-IR wavelengths used in this study. The 3-$\sigma$ upper limits of the magnitude for {\it HST/WC3} F555W was found as 27.2 mag while no upper limits were found for other optical filters since they were well below the detection limit of the {\it WFC3} detector. The counterpart has an average Vega magnitude of $\sim$ 20.1 in NIR images and $\sim$ 14.9 in mid-IR images. The absolute magnitude F200W filter ($\sim$ K-band) was derived as -10.5 mag. This value is different from the average absolute K-band value of RSGs in Small Magellanic Cloud (e.g., \citealp{2009AJ....138.1003B}).\\

X-ray transient AGNs and galaxies, similar to ULXs, can be very bright or even too faint to be observed and they exhibit irregular variations in the long-term X-ray light curves. In this case, the possibility of ULX-3 as an AGN should also be addressed. X-ray transient AGNs can also exhibit variability in the optical band, but the variability of ULX-3 in these bands is not clear due to the lack of both IR and optical data. F$_{X}$/F$_{optik}$ ratios may be used to distinguish clusters of galaxies, BL Lac, XRBs, and normal galaxies as well as AGNs \citep{1991ApJS...76..813S}. To classify counterpart ULX-3, this ratio was derived using Equation \ref{fx} given by \cite{1982ApJ...253..504M}.

\begin{equation}\label{fx}
 log(F_X/F_{optik}) = log(F_X) + m_V/2.5 + 5.37
\end{equation}
where the F$_{X}$ is observed absorbed flux obtained from a simple {\it power-law} in the 0.3–3.5 keV energy range and m$_{V}$ is the V-band magnitude. Since ULX-3 was not observed {\it HST/WFC3} F555W ($\sim$ V-band) image 3-$\sigma$ upper limit Vega magnitude was derived as 27.2. The absorbed fluxes were calculated from X-ray observations when the source was detected. F$_{X}$/F$_{optik}$ range was found as 3.1-3.8 which excludes ULX-3 as an AGN or type of other galaxy. Due to the feature of the X-ray variability of ULX-3, simultaneous X-ray and V-band observations are needed for this ratio. Moreover, counterpart ULX-3 was compared with the V-I colors of 144 AGNs (F555W-F814W in this study) within the LMC \citep{2012ApJ...746...27K}. In their study, the range V-I colors of AGNs is mostly 0.9 mag, while for counterpart ULX-3 is $\sim$ 1.5 mag, and also AGNs are brighter than ULX-3 in V-band magnitude (V-band = $\sim$ 19).\\

To constrain the age of the possible donor star CMD F3000M vs F200W-F300M for NIR emissions was plotted by assuming the NIR emission is dominated by the counterpart and also, and it is also associated with neighboring star candidates. Based on CMD, the age of the counterpart was estimated as 4-12 Myr. If the age range of the possible donor is as given, the masses corresponding to these ages should be 18-30 M$\odot$ and these results are consistent with RSGs \citep{2021ApJ...907...18R}. Moreover, the K-band magnitude is consistent with the extra-galactic RSG ULXs reported in previous studies. In the case of another CMD M${F360M}$ vs F360M-F770W, ULX is well-consistent with supergiant B(e) type stars. The conclusion from the CMDs is that ULX-3 is more likely to be an SgB(e) star and this feature is compatible with the results of X-ray analysis. Because of the dynamical effects on the B(e) star disk, some HMXBs with B(e) star donors can occasionally produce high mass transfer rates for short-term periods. These systems possess the distinctive quality of undergoing recurrent transitions between a typical X-ray binary state and a ULX state \citep{2023NewAR..9601672K}.\\

Although {\it blackbody} radiation is expected from the donor star the {\it blackbody} temperatures (200 K and 1000 K) obtained from the SED of the counterpart ULX-3 are too low to be RSG \citep{2005ApJ...628..973L}. Therefore, ruled out the possibility that the donor star of ULX-3 is an RSG and red giant. As seen in Figure \ref{F:SED}, NIR emission is represented by a {\it blackbody} model with a temperature of 1000 K, while mid-IR emission is represented by a {\it blackbody} model with a temperature of 200 K. However, if indeed these temperatures come from the possible donor star, it can not be an SgB(e) star, since the temperature of such these stars is usually around $\geq$10000 K (e.g. \citealp{2021A&A...647A..28K}). Another scenario for IR emission of ULX-3 that variable jet activity such as for Holmberg IX X-1 which exhibited the dramatic variable behavior of the mid-IR \citep{2016ApJ...831...88D,2019ApJ...878...71L} but, since there is no such evidence for ULX-3, this possibility was ruled out.\\

Since the emissions at IR wavelengths are less affected by dust extinction the main reason why the counterpart of ULX-3 could not be detected in optical and UV except the red filter (F814W) could be due to circumbinary disk/dust. Moreover, a large fraction of the incident UV/optical emission could be absorbed by the possible circumbinary/dust and re-emitted at NIR wavelengths. In the case of thermal emission from a circumstellar/binary disk of the SgB(e) star, we would expect thermal emission with a temperature of a few hundred K high up to 1000 K. Following a similar approach \cite{2017ApJ...838L..17L,2019ApJ...878...71L}, the equilibrium temperature radius provides an estimate of the lower limit or inner radius of
the surrounding warm dust, R${eq}$ derived by Equation \ref{E6}:
\begin{equation}\label{E6}
R_{eq} = \frac{Q_{abs}L_{IR}}{Q_{e}16\pi \sigma T^{4}}
\end{equation}
where {\it Q$_{abs}$} and {\it Q$_{e}$} are the grain absorption and emission Planck mean cross sections, {\it $\sigma$} is Stefan-Boltzmann constant, L$_{IR}$ is the luminosity for a heating source SgB(e) 20000 K. In the case of radiative dust, heating is dominated by the stellar component, {\it L$_{\star}$} = {\it L$_{IR}$} = 2 $\times$ 10$^{4}$ L${\odot}$. The value of {\it Q$_{abs}$}/{\it Q$_{e}$} are $\sim$ 20 and 100 for NIR and mid-IR, respectively. {\it T} is temperatures obtained from SED of ULX-3. The R${eq}$ were derived as $\sim$ 50 and 2800 au (astronomical unit) for NIR and mid-IR emissions. \\

According to the conclusion from the two CMDs, that is if we accept that the donor star of ULX-3 is an SgB(e), we can understand further insight into the circumstellar dust geometry such as estimating the binary separation, {\it a}. In the case of a possible SgB(e) donor star, we would expect the bright X-ray flux from ULX-3 to transfer a significant amount of mass from the donor star to the accretor, possibly via Roche Lobe overflow. Using a mass of 25 M$\odot$ and a radius {\it R$_{\star}$} = 27 R$\odot$ for the well-known SgB(e) source IGR J16318-4848 \citep{2004ApJ...616..469F,2008A&A...484..801R}, a rough estimate of the binary separation value was derived as {\it a}= $\sim$ 0.3 ua from the Roche lobe radius {\it R$_{L}$} \citep{1983ApJ...268..368E} Equation \ref{E7}. 
\begin{equation}\label{E7}
\frac{R_{L} }{a} = \frac{0.49q^{2/3}}{0.6q^{2/3}+log(1+q^{1/3})}
\end{equation}
In this case, the mass of the compact object was assumed to be 25 M$\odot$ and {\it R$_{L}$} = {\it R$_{\star}$} was accepted. Given that {\it a} < R$_{eq}$ $\sim$ 50-2800 20 au, the dust is likely in a circumbinary disk. In this case, of the two {\it blackbody} models fitted to the SED of ULX-3, the higher temperature (1000 K) represents emission from the inner circumbinary disk, while the lower temperature (200 K) probably originates from the outer part of the circumbinary disk. The SgB(e) stars have significant mass loss rates. If we assume that the material of the circumbinary disk is optically thin we can estimate the lower limit mass of the disk/dust and then may estimate the mass loss rate (e.g. \citealp{2019ApJ...878...71L}). However, avoiding the calculation of these parameters as powerful assumptions are required for these calculations, therefore, we need spectral and interferometer observations. As a result, typical SgB(e) are surrounded by circumbinary regions, but since the true colors of ULX-3 are unknown due to the circumbinary disk or hot dust, it is not clear whether its counterpart is an SgB(e). In other words, the CMD results are not very reliable since the observed IR emission of ULX-3 comes mostly from the circumbinary disk/dust.

\section{Conclusions and summary}\label{sec:5}

All the available X-ray data were analyzed for the new transient source, ULX-3, in galaxy NGC 4254. The X-ray data were taken from the {\it Chandra} and {\it Swift-XRT} and the Optical and IR data were taken from {\it HST} and {\it JWST} archives. The two main goals of the analysis of the ULX-3 were to determine the nature of X-ray emission and identify the optical/NIR/mid-IR counterparts. Limited X-ray spectral analysis was done by deploying simple absorbed spectral models to ascertain and describe the main spectral features of the ULX-3. Additional spectral analysis of the X-ray data included the extraction of hardness ratios to identify possible spectral transitions of the ULX-3. Precise astrometric calculations were performed to determine the possible donor star of ULX-3. CMDs and SED were plotted to determine the type of the possible donor star. Two {\it blackbody} temperatures were determined, which could possibly come from the inner and outer regions of the circumbinary disk. The main findings from this study are summarized as follows:\\

A new transient ULX candidate (ULX-3) with reaching peak luminosity of 4 $\times$ 10$^{39}$ erg s$^{-1}$ was identified from \textit{Swift/XRT} X-ray data in NGC 4254. According to the X-ray hardness ratios, ULX-3 exhibits very hard-to-soft transitions as seen in some HMXBs with Be-star donors. As a typical transient ULX, ULX-3 varies by more than two orders of magnitude. Resulting in precise astrometric calculations revealed that ULX-3 has unique optical, NIR, and mid-IR counterparts within a 0.55 arcsec error radius. From the SED, two {\it blackbody} temperatures (1000 K and 200 K) were defined for the IR counterpart which can be attributed to the heating of dust in a circumbinary disk.

\section*{Acknowledgements}

This paper was supported by the Scientific and Technological Research Council of Turkey (TÜBİTAK) through project number 122C183. I would like to thank the Referee for his helpful comments and suggestions, which helped to clarify some issues. I would like to thank A. Akyuz for her valuable suggestions. I would like to thank my heartfelt appreciation to my wife Semiha Allak for her patience during the entire process of writing this paper. Finally, I would also like to thank my little son Atlas Allak who is the sweetest source of my motivation.

\section*{Data Availability}

The scientific results reported in this article are based on archival observations made by the James Webb Space Telescope and Hubble Space Telescope, and obtained from the data archive at the Space Telescope Science Institute\footnote{https://mast.stsci.edu/portal/Mashup/Clients/Mast/Portal.html}. This work has also made use of observations made with the {\it Chandra}\footnote{https://cda.harvard.edu/chaser/} and {\it Swift/XRT}\footnote{https://swift.gsfc.nasa.gov/about\_swift/xrt\_desc.html} X-ray Observatories.


\bibliographystyle{mnras}
\bibliography{ngc4254} 

\bsp	
\label{lastpage}
\end{document}